%% file: ms.tex
[1994/12/01]
\documentclass[manuscript]{article}

\title{A Turing Incomputable Coloring Function}
\author{Michael Stephen Fiske}

\usepackage{amsmath,amsthm,amsfonts,amssymb,mathtools,braket}
\usepackage{graphics}

\chardef\bslash=`\\ 

\newtheorem{thm}{Theorem}[section]
\newtheorem{Corollary}[thm]{Corollary}
\newtheorem{Lemma}[thm]{Lemma}

\theoremstyle{definition}
\newtheorem{defi}{Definition}[section]

\theoremstyle{remark}
\newtheorem{Remark}{Remark}[section]

\newtheorem{example}{Example}

\newcommand{\eval}[2][\right]{\relax
  \ifx#1\right\relax \left.\fi#2#1\rvert}

\begin{document}

\maketitle

\begin{abstract}
   This paper describes a sequence of natural numbers that grows faster than any Turing 
   computable function.   This sequence is generated from a version of the tiling problem, called a coloring system.   
   In our proof that generates the sequence, we use the notions of a chain and an unbounded sequence property, 
   which resemble the methods of point set topology.     
   From this sequence,  we define a Turing  incomputable coloring function.  
\end{abstract}


\input{coloring_numbers_arxiv.tex}

\end{document}

%% file: coloring_numbers_arxiv.tex


\section{Introduction}\label{C:one}

In  \cite{berger:mib}, Berger showed that the domino problem was unsolvable.    This is well-known as the tiling problem \cite{wang:mib}.   
We define the tiling problem in terms of a coloring system.  

\smallskip

\begin{defi}\label{CS:1}     \hskip 1pc      {\it  Coloring System}

\noindent  A coloring system is a quadruple  $(C, a, H, V)$ satisfying three conditions: 

\smallskip 

(a)  \hskip 0.5pc   $C$ is a finite set.    

\smallskip

(b)  \hskip 0.5pc   The color $a$ lies in set $C$.   

\smallskip

(c)  \hskip 0.5pc    A horizontal set $H \subseteq C \times C$ and a vertical set $V \subseteq C \times C$.

\end{defi}

\smallskip 

$C$ is called a color set.  $H$ and $V$ are called coloring rules.  
The elements of $C$ can be represented as the first $n-1$ counting numbers and $0$. That is, $C = \{0, 1, 2, \dots, n-1 \}$, and 
every $C$ is a proper subset of $\mathbb{N} = \{0, 1, 2, 3, 4, \dots \}$ (natural numbers).

\smallskip

\begin{defi}\label{CS:2}     \hskip 1pc      {\it  Acceptable Coloring}

\noindent  An  acceptable coloring of coloring system $(C, a, H, V)$, is a function                        
$f :  \mathbb{N} \times \mathbb{N}  \rightarrow  C$  satisfying three conditions:   

\smallskip

(1)   \hskip 0.5pc    $f(0, 0) = a$. 

\smallskip

(2)   \hskip 0.5pc    For all $m, n  \in  \mathbb{N}$,   $\big{(}  f(m, n),$  $f(m+1, n) \big{)}   \in H$. 

\smallskip

(3)    \hskip 0.5pc   For all $m, n  \in  \mathbb{N}$,   $\big{(}  f(m, n),$  $f(m, n+1)  \big{)}  \in V$.

\end{defi}

\smallskip

\noindent  A tile is an element $(m, n)$ of $\mathbb{N} \times \mathbb{N}$.   
If an acceptable coloring $f$ exists, $f$ determines the color of each tile,   
according to the coloring rules $H$ and $V$.

\smallskip

\begin{example}  \hskip 1pc      {\it  Acceptable Coloring}

\noindent We describe a non-trivial coloring system that has an acceptable coloring.
Set $C = \{0, 1, 2, 3, 4, 5, 6, 7, 8, 9, 10, 11, 12 \}$.   $0$ is red.  $1$ is blue.  $2$ is forest green.  
$3$ is purple.  $4$ is yellow.  $5$ is pink.  $6$ is aqua.    $7$ is  grey.  $8$ is teal.  $9$ is lime green.   
$10$ is brown.   $11$ is candy green. $12$ is orange.  Origin color $a = 1$  is blue.

\medskip

\noindent    $H = \{  (0, 1), (0, 4),  (1, 2), (1, 5), (2, 0), (2, 3), (3, 2), (3, 5), (4, 0), (4, 3), (5, 1), (5, 4),$

\hskip 0.9pc  $(6, 6), (6, 9), (6, 10), (7, 7), (7, 8), (7, 11), (7, 12), (8, 6), (8, 9), (8, 10), (9, 7),   $

\hskip 0.9pc  $(9, 8), (9, 11), (9, 12), (10, 6), (10, 9), (10, 10), (11, 7), (11, 8), (11, 11), (11, 12),  $

\hskip 0.9pc  $ (12, 6), (12, 9), (12, 10)   \}$.

\medskip

\noindent   $V = \{  (0, 2), (0, 4),  (0, 5), (0, 8), (0, 9), (0, 12), (1, 2), (1, 4), (1, 5), (1, 8), (1, 9), (1, 12),  $   

\hskip 0.9pc   $(2, 2), (2, 4), (2, 5), (2, 8), (2, 9), (2, 12), (3, 10), (3, 11), (4, 10), (4, 11), (5, 10),   $

\hskip 0.9pc   $(5, 11), (6, 0), (6, 1), (6, 3), (7, 0), (7, 1), (7, 3), (8, 0), (8, 1), (8, 3), (9, 2), (9, 4),   $
 
\hskip  0.9pc  $(9, 5), (9, 8), (9, 9), (9, 12), (10, 2), (10, 4), (10, 5), (10, 8), (10, 9), (10, 12),  $

\hskip  0.9pc  $(11, 2), (11, 4), (11, 5), (11, 8), (11, 9), (11, 12), (12, 6), (12, 7)  \}$.

\medskip

\noindent    Below is a partial coloring of the tiles  $S = \{ (x, y) \in  \mathbb{N} \times \mathbb{N}:   0 \le  x + y \le 9  \}$.

\begin{center}

          $ \begin{matrix}
                         8 \\
                        9    &   11  \\
                        1    &    5      &    4    \\
                        8    &   10   &    9  &  8    \\
                        0    &    4     &    0  &  1    &  2     \\
                        8    &    9     &    7  &  8    &  9   &  8  \\
                        9    &   11  &  12  & 10  &  9   & 11  & 12  \\
                        0    &    4    &  0    &  4    &  0   &  4   & 0 & 4\\
                        8    &    9    &   8   &  9    &  8   &  9   &  8  & 9   &  8   \\
                        1    &    2    &   0   &  1    &  2   &  0   &  1  &  2  &  0  &  1  \\
          \end{matrix}$

\end{center}

\end{example}

\smallskip

\begin{defi}\label{CS:3}     \hskip 1pc      {\it  Partial Coloring}

\noindent  A partial coloring is a function $g:  S  \subset  \mathbb{N} \times \mathbb{N}   \rightarrow  C$  satisfying the same conditions as an acceptable coloring except $g$ is restricted to $S$:   

\smallskip 

(a)  \hskip 0.5pc  $g(0, 0) = a$.

\smallskip

(b)  \hskip 0.5pc  For all $(m, n)  \in  S$,   $\big{(}  g(m, n),$  $g(m+1, n) \big{)}   \in   H$. 

\smallskip

(c)  \hskip 0.5pc   For all $(m, n)  \in  S$,   $\big{(}  g(m, n),$   $g(m, n+1) \big{)}   \in  V$. 

\end{defi}

\smallskip

The {\it coloring problem} is the following:    
given a particular coloring system $(C, a, H, V)$ determine whether there exists an acceptable coloring $f$ that satisfies   $(C, a, H, V)$.
The coloring problem is unsolvable by the class of computing machines equivalent to a Turing machine, as discussed in 
\cite{turing:mib}, \cite{sturgis:mib}, \cite{berger:mib}, and \cite{davis:mib}.

A diagonal map and inverse diagonal map are defined so that acceptable color sequences and acceptable coloring spaces can be defined.

\begin{defi}\label{CS:4}     \hskip 1pc      {\it  Diagonal Map}

\noindent  Define diagonal map  $D:  \mathbb{N} \times \mathbb{N} \rightarrow \mathbb{N}$, where  $D(x, y) =  {\frac {1}  {2}}   (x+y) (x + y+ 1)   + x$.

\end{defi}

\smallskip

\noindent  Observe that $D(0, 0) = 0$,  $D(0, 1) = 1$,  $D(1, 0) = 2$, $D(0, 2) = 3$, $D(1, 1) = 4$, $D(2, 0) = 5$,   . . .   
In other words, $D$ diagonally counts the tiles.


\begin{center}
            $ \begin{matrix}
                       10   &      \\
                        6    &    11    \\
                        3    &    7  &  12   \\
                        1    &    4  &   8   &   13    \\
                        0    &    2  &   5   &  9 &  14    \\
              \end{matrix}$
\end{center}


The inverse diagonal map is defined $D^{-1}(n) = (X(n), Y(n) )$ where $T(n) = {\frac 1 2} n (n+1)$.  
$L(n) = max\{ k \in \mathbb{N}: T(k) \le n \}$.  \hskip 0.5pc
$X(n) = n - T \circ L(n)$.   $Y(n) = L(n) - X(n)$. 
For example, $D^{-1}(5) = ( X(5), Y(5) ) = (2, 0)$ so  5 corresponds to tile $(2, 0)$.

\smallskip

A finite sequence  $(a_0,  a_1,  a_2, \dots, a_n)$  of colors, called a color sequence,  chosen from $C$ defines
a coloring of the first  $n+1$  tiles using the inverse diagonal map.   $D^{-1}(0) = (0, 0)$, so tile $(0, 0)$  is colored $a_0$.  
Tile $D^{-1}(k)$  is colored  $a_k$.

\smallskip

\begin{defi}\label{CS:5}     \hskip 1pc      {\it  Acceptable Color Sequences}

\noindent      Define the set  $S = \{ (j, k) :   j ,  k \ge 0$  \verb|and| $D(j, k) \le n \}$.   Define the partial coloring  
$g:  S \rightarrow C$  where  $g(j, k)   =  a_{D(j, k)}$.   Color sequence  $(a_0,  a_1,  a_2, \dots, a_n)$
satisfies coloring system $(C, a, H, V)$  if partial coloring $g$ satisfies  $(C, a, H, V)$.   Sometimes  
$(a_0,  a_1,  a_2, \dots, a_n)$  is called an acceptable coloring sequence of  $(C, a, H, V)$. 

\end{defi}

\smallskip

\begin{defi}\label{CS:6}     \hskip 1pc      {\it  Acceptable Coloring Space}

\noindent      Define the acceptable coloring space  $ {\mathfrak A}(C, a, H, V) = \{ (a_0,  a_1,  a_2, \dots, a_n): $  for each  $k$ such that   
$0 \le k \le n$, tile $D^{-1}(k)$ is colored $a_k$ where  $a_k  \in C$, and color sequence $(a_0,  a_1,  a_2, \dots, a_n)$  satisfies  $(C, a, H, V)$ $\}$.

\end{defi}


 \section{An Incomputable Coloring Function}\label{A:two}

The main theorem is the following.    If no acceptable coloring exists for coloring system $(C, a, H, V)$,   then there exists a number 
$M$ such that all sequences in  ${\mathfrak A}(C, a, H, V)$  have length less than  $M$.     
Consequently, for a fixed number of colors $C = \{c_1, c_2, \dots c_n\}$
the corollary constructs an incomputable bound  $M_n$ over all spaces  ${\mathfrak A}(C, a, H, V)$ 
such that no acceptable coloring exists  for  $(C, a, H, V)$.    
A Turing  incomputable function  $\mu$  is defined as $\mu(n) = M_n$.

The rest of this section works toward proving theorem \ref{thm_inf_sequence_coloring} and corollary \ref{corollary_incomputable_function}.  
First, some preliminary definitions and remarks are developed.  
Let $S$ be a finite or infinite sequence (of colors) of elements of $C$.  The  $k$th element of  $S$ is denoted as $S_k$.   
The length of sequence  $P$,  
denoted  $|P|$, is the number of elements in the sequence.   
For  example, if  $P = (c$, $c$,  $c)$, then  $|P| = 3$.  
If  $P$  is an infinite sequence, then $|P|  =  \infty$.

\smallskip

\begin{defi}\label{A:1}     \hskip 1pc

 \noindent  The sequence $S$ contains sequence $P$, denoted $P \subseteq S$,  if  $|S|  \ge  |P|$  and  $S_k  =  P_k$  for every  $k$ where  $0  \le  k  \le   |P|  -  1$.

 \end{defi}

 
\begin{defi}\label{A:2}     \hskip 1pc      {\it  Chain}

\noindent    A sequence of sequences  $(S_0, S_1, S_2, S_3,  \dots)$  is a chain if  $S_0 \subseteq S_1  \subseteq S_2  \subseteq S_3  \subseteq  \dots$
where the chain can be a finite or infinite number of sequences.   

\end{defi}


\noindent  Let $S = (s_0,  s_1,  s_2, \dots, s_m)$.  Define $S^k = (s_0,  s_1,  s_2,$ 
$\dots, s_k)$, where  $0 \le k  \le  m$.

\smallskip

\begin{Remark}\label{C:subsequence}    \hskip 1pc    
 
\noindent    If  $S = (s_0,  s_1,  s_2, \dots, s_m)  \in  {\mathfrak A} (C, a, H, V)$, then each sequence  $S^0,  S^1, S^2, \dots , S^m$  lies  in   ${\mathfrak A} (C, a, H, V)$. 

\end{Remark}

\proof   This follows immediately from the definition of  ${\mathfrak A} (C, a, H, V)$.    

\smallskip

\begin{defi}\label{defi_unbounded_seq_property}     \hskip 1pc   {\it Unbounded sequence property}  \hskip 0.5pc   $(s_0,  s_1, ..., s_k)$

\noindent     A sequence  $(s_0,  s_1, ..., s_k)  \in {\mathfrak A} (C, a, H, V)$  has the unbounded sequence property if for 
each  $M  \ge  k$, there is a sequence,  $(a_0,  a_1,  \dots, a_M)$, with length  $M+1$,  such that
  $(a_0,  a_1,   \dots, a_M)  \in   {\mathfrak A} (C, a, H, V)$, and satisfies    $(s_0,  s_1, \dots, s_k)   \subseteq  (a_0,  a_1,   \dots, a_M)$.  In 
other words, $s_i = a_i$   for every  $i$  such that  $0  \le  i  \le  k$. 

\end{defi}

\smallskip  

\begin{defi}\label{defi_unbounded_seq_property_A}     \hskip 1pc   {\it Unbounded sequence property} \hskip 0.5pc  ${\mathfrak A} (C, a, H, V)$

\noindent     Set  ${\mathfrak A} (C, a, H, V)$  has the unbounded sequence property if for  
every natural number  $M$, there is a sequence $S$ of length $M+1$   such that $S \in   {\mathfrak A} (C, a, H, V)$.

\end{defi}

\smallskip  

\begin{Lemma}\label{lemma_c_infinity}   \hskip 1pc   

\noindent     If sequence  $(s_0,  s_1,  s_2, \dots,  s_k)  \in {\mathfrak A} (C, a, H, V)$ has the unbounded sequence property, 
then there exists a color $c_{\infty} \in C$ such that sequence 
$(s_0,  s_1,  s_2, \dots, s_k, c_{\infty})$   has the unbounded sequence property.   

\end{Lemma}

\proof      Let   $(s_0,  s_1,  s_2, \dots,  s_k)  \in {\mathfrak A} (C, a, H, V)$  with the unbounded sequence property.  Set  
$M_1 = k + 1$.  From the unbounded sequence property, there exists  $B^1 = (a_0,  a_1,  a_2,  \dots, a_{k+1})$  
$\in   {\mathfrak A} (C, a, H, V)$  such that $a_i = s_i$   whenever $0  \le  i  \le  k$.   From induction and the unbounded sequence property, for each 
natural number $q$ where $M_q = k + q$, there exists $B^q \in  {\mathfrak A} (C, a, H, V)$  with length  $k + q + 1$
such that  $(s_0,  s_1,  s_2, \dots,  s_k)    \subseteq   B^q$.  Since each  $B^q$  exists, define

$$W =  \underset{q=1}  { \overset{\infty}  \cup  } \{ B^q \in  {\mathfrak A} (C, a, H, V):  (s_0,  s_1,  s_2, \dots,  s_k)  \subseteq B^q \}.$$

Let  $(B^q)_{k+1}$ denote the $k+1$th color of sequence $B^q$.   Consider the infinite sequence of colors $Z = ( (B^1)_{k+1}, (B^2)_{k+1},  (B^3)_{k+1}, \dots )$.  Since the color set $C$
is finite, there is at least one color, called $c_{\infty}$, that occurs infinitely often in $Z$.  The existence of $c_{\infty}$ and the following claim completes the proof.

\bigskip

\noindent  Claim:  \hskip 0.5pc  $(s_0,  s_1,  s_2, \dots,  s_k, c_{\infty} )$ has the unbounded sequence property.   

\smallskip

\noindent  Proof of the claim.   Let  $M$ be a natural number such that $M \ge k+1$.  
Since $c_{\infty}$ occurs infinitely often in $Z$, there is a sequence $B^p \in W$ where $(B^p)_{k+1} = c_{\infty}$ 
and $p \ge M$.    This implies that  $(s_0,  s_1,  s_2, \dots,  s_k, c_{\infty} )   \subseteq  B^p$, so the claim is proven.

\smallskip

\begin{Remark}\label{rem_singleton_unbounded_seq_property}    \hskip 1pc    
 
\noindent   If ${\mathfrak A} (C, a, H, V)$ has the unbounded sequence property,  then the sequence $(a)$ has the unbounded sequence property.  

\end{Remark}

\proof    Let  $M$ be a natural number.   Since ${\mathfrak A} (C, a, H, V)$ has the unbounded sequence property, there exists 
$(a_0,  a_1,  a_2, \dots,  a_M)   \in   {\mathfrak A} (C, a, H, V)$.   By the definition of an acceptable coloring, any element of ${\mathfrak A} (C, a, H, V)$ must begin with $a$.   
Thus $a_0 = a$, so  $(a) \subseteq  (a_0,  a_1,  a_2, \dots,  a_M)$.

\smallskip

\begin{Lemma}\label{lemma_unbounded_seq_property_chain}   \hskip 1pc   

\noindent     If  ${\mathfrak A} (C, a, H, V)$ has the unbounded sequence property, then there exists an infinite chain of sequences 
$(a_0)  \subseteq  (a_0, a_1) \subseteq (a_0, a_1, a_2) \subseteq \dots  (a_0, a_1, a_2, \dots, a_k) \subseteq \dots $ such that  
$(a_0, a_1, a_2,$   $\dots, a_k)$   $\in$  ${\mathfrak A} (C, a, H, V)$ for every natural number $k$.  

\end{Lemma}

\proof   Using induction, we prove that for every $k$, we can construct a sequence        
$(a_0, a_1, a_2,$  $\dots, a_k)$  that has the unbounded sequence property.  Then the lemma follows 
from this.  The base case,  $m = 0$, holds by setting  $a_0 = a$ and applying remark \ref{rem_singleton_unbounded_seq_property} to the 
sequence $(a_0)$.    
 
For the inductive step, $m = k$, suppose we have constructed $(a_0, a_1, a_2,$  $\dots, a_k)$ with the 
unbounded sequence property.  Lemma  \ref{lemma_c_infinity}  implies there is a color $a_{k+1}$ such that           
 $(a_0, a_1, a_2, \dots, a_k, a_{k+1})$  has the unbounded sequence property.  Thus, our induction proof 
shows that an infinite chain of sequences  
$(a_0)  \subseteq  (a_0,  a_1)  \subseteq   (a_0,  a_1,  a_2)  \subseteq   \dots  (a_0,  a_1,  a_2, \dots, a_k)  \subseteq   \dots$  
can be constructed such that each sequence is an element of   ${\mathfrak A} (C, a, H, V)$.

\smallskip

\begin{thm}\label{thm_inf_sequence_coloring}    \hskip 1pc  

 \noindent  If for any natural number $M$, there exists a sequence, 
$S$, of length greater than $M$ such that  $S \in {\mathfrak A} (C, a, H, V)$, then there exists an infinite 
sequence $(s_0,  s_1 ,  s_2   \dots) \in  {\mathfrak A} (C, a, H, V)$.  In other words, there exists an acceptable 
coloring which satisfies the coloring system $(C, a, H, V)$.

\end{thm}

\proof\        The hypothesis means that $ {\mathfrak A} (C, a, H, V)$  has  the unbounded sequence property.  
Lemma \ref{lemma_unbounded_seq_property_chain} implies there exists an infinite chain of sequences 
$(a_0)  \subseteq  (a_0,  a_1)   \subseteq  (a_0,  a_1 ,  a_2)   \subseteq  \dots  (a_0,  a_1,  a_2, \dots, a_k)   \subseteq  \dots$  
such that  $(a_0,  a_1, \dots, a_k)  \in     {\mathfrak A} (C, a, H, V)$  for every natural number $k$.  From the infinite chain, define the infinite 
sequence   $(a_0,  a_1,  a_2,$  $\dots, a_k, \dots)$.  For every $k$,  $(a_0,  a_1,  a_2, \dots, a_k)  \in 
 {\mathfrak A}(C, a, H, V)$.  This means that the colors assigned to the tiles represented by infinite 
sequence $(a_0,  a_1,  a_2,$  $\dots,$  $a_k, \dots)$ satisfy coloring system $(C, a, H, V)$.  Thus,              
$(a_0,  a_1,  a_2, \dots, a_k, \dots)  \in  {\mathfrak A} (C, a, H, V)$,  so  $(a_0,  a_1,  a_2,$ $\dots, a_k, \dots)$  
is an acceptable coloring of  $(C, a, H, V)$.

\smallskip

\begin{defi}  \hskip 1pc  {\it Isomorphic Coloring Systems} 

\noindent  Coloring system $(C, a_1, H_1, V_1)$ is isomorphic to $(C, a_2, H_2, V_2)$ 
if there exists a bijective map $\phi: C \rightarrow C$ such that all three conditions hold:

\smallskip 

(1)  \hskip 0.5pc  $\phi(a_1) = a_2$.

\smallskip  

(2)  \hskip 0.5pc  Pair $(m, n)$ lies in $H_1$ if and only if  $\big{(} \phi(m), \phi(n) \big{)}$ lies in $H_2$.  

\smallskip 

(3)  \hskip 0.5pc  Pair $(m, n)$ lies in $V_1$ if and only if  $\big{(} \phi(m), \phi(n) \big{)}$ lies in $V_2$. 

\end{defi}

\smallskip  

\begin{Corollary}\label{corollary_incomputable_function}    \hskip 1pc  

 \noindent   Suppose  $|C| = n$.  There exists an upper bound  $M_n$  such that for every coloring system  $(C, a, H, V)$  where no acceptable coloring exists, 
then all sequences from the space $  {\mathfrak A}(C, a, H, V)$  have length less than $M_n$. 
Moreover, $\mu(n) = M_n$ is a Turing incomputable function.

\end{Corollary}

\proof\     Since $C$ has  $n$ elements, there are a finite number of coloring systems.  In particular, there are $n$ choices for $a$.
There are $n^2$ ordered pairs $(m, n)$, so  each pair $(m, n)$ is in $H$ or not in $H$.  
Similarly, each pair $(m, n)$ is in $V$ or not in $V$. 
Thus, there are $n \big{(} 2^{n^2} \big{)} \big{(} 2^{n^2} \big{)}$  coloring systems,  where some can be isomorphic.  
Index the coloring systems for which no acceptable coloring exists, with the numbers $\{1, 2, 3, \dots, q\}$, where    
$q < n \big{(} 2^{n^2} \big{)} \big{(} 2^{n^2} \big{)}$.   For each coloring system $(C, a, H, V)$ where no acceptable 
coloring exists, indexed as $k$, theorem \ref{thm_inf_sequence_coloring} implies there is a number $U_k$ so that all 
sequences in $  {\mathfrak A}(C, a, H, V)$ have length less than $U_k$.  Set  $M_n  =$  \verb|max|$\{U_1, \dots, U_q \}$.

For each $n$ ($|C| = n$), function  $\mu$  is defined as $\mu(n) = M_n$.   If  $\mu$ could be computed by a Turing machine, 
then the coloring problem would be Turing machine solvable by finding $(C, a, H, V)$ acceptable sequences up to $M_n$.  
From the previous paragraph,  if an acceptable sequence has length $M_n$ or greater, 
then $(C, a, H, V)$ has an acceptable coloring.   This contradicts that the coloring problem has been shown to be unsolvable by a Turing machine.   
Hence, $\mu$ is a Turing incomputable function.


%% file: ms.bbl
\begin{thebibliography}{}
  
    \bibitem[Ber66]{berger:mib}
    Robert Berger.
    \newblock The undecidability of the domino problem.
    \newblock{\em Memoirs of the American Mathematical Society}, Number 66, 1966. 
    
    \smallskip
     
    \bibitem[Dav82]{davis:mib}
    Martin Davis. 
    \newblock     {\em  Computability and Unsolvability}.     Dover Publications,   New York, 1982.
    
    \smallskip
   



    \bibitem[Stu63]{sturgis:mib}
    H. E.  Sturgis,   and  J. C.  Shepherdson.  
    \newblock    Computability of Recursive Functions.
    \newblock   {\em Journal Assoc. Computing Machines},  Vol. 10,  217-255, 1963.            
    

    \smallskip       


    \bibitem[Tur36]{turing:mib}
    Alan M. Turing.
    \newblock    On computable numbers, with an application to the Entscheidungsproblem.
    \newblock    {\em Proc. London Math. Soc.}  ser. 2   {\bf  42} (Parts 3 and 4), 230-265,  1936; [Turing, 1937a] A correction,   ibid. {\bf 43},  544-546,  1937.      
     


    \smallskip 

    
    \bibitem[Wan61]{wang:mib} Hao Wang.
    \newblock      Proving theorems by pattern recognition II. 
    \newblock      {\em Bell Systems Technical Journal}.  Vol. 40, Issue 1, 1–41, January 1961.

    
      
\end{thebibliography}
